\documentclass[11pt]{article}

\usepackage[preprint]{acl}

\usepackage{times}
\usepackage{latexsym}
\usepackage{booktabs} 
\usepackage{multirow} 
\usepackage{amsmath}
\usepackage[most]{tcolorbox}
\usepackage{xcolor}
\usepackage{enumitem}
\usepackage{listings}
\usepackage{hyperref}
\usepackage{subcaption}

\newtcolorbox{promptbox}[1]{
    enhanced,
    breakable,
    colback=blue!3,
    colframe=blue!55!black,
    colbacktitle=blue!55!black,
    coltitle=white,
    fonttitle=\bfseries,
    title=#1,
    boxrule=0.8pt,
    arc=1.8mm,
    outer arc=1.8mm,
    left=2mm,
    right=2mm,
    top=1.5mm,
    bottom=1.5mm,
    before skip=6pt,
    after skip=6pt,
    fontupper=\small
}

\lstdefinestyle{jsonstyle}{
    basicstyle=\ttfamily\scriptsize,
    breaklines=true,
    columns=fullflexible,
    keepspaces=true,
    showstringspaces=false,
    frame=none,
    xleftmargin=0pt,
    aboveskip=4pt,
    belowskip=4pt
}

\usepackage[T1]{fontenc}

\usepackage[utf8]{inputenc}

\usepackage{microtype}

\usepackage{inconsolata}

\usepackage{graphicx}

\title{Learning from Real Stuttering: Synthetic Mandarin Stuttering Speech for Data Augmentation in Speech Recognition}

\author{
  \textbf{Yao Lu, Jiaming Zhou, Zhongyuan Fu, Yong Qin\textsuperscript{*}} \\
  College of Computer Science, Nankai University \\
  \small{
    \textbf{Correspondence:} 
    \href{mailto:nku_luyao@mail.nankai.edu.cn}{nku\_luyao@mail.nankai.edu.cn},
    \href{mailto:qinyong@nankai.edu.cn}{qinyong@nankai.edu.cn}
  } \\
}

\begin{document}
\maketitle
\begin{abstract}
Mandarin stuttered speech remains under-resourced, limiting the development of robust speech processing systems. We present a Mandarin stuttering speech synthesis and data augmentation framework that adapts both autoregressive and non-autoregressive TTS models using naturally stuttered speech. Specifically, we adapt CosyVoice2 and MaskGCT to support controllable generation of specific stuttering events while also modeling broader utterance-level characteristics of stuttered speech. Using the adapted models, we construct synthetic data resources for downstream augmentation. Evaluations show that the generated speech closely approaches real stuttered speech. Extensive ASR experiments demonstrate consistent augmentation gains and further observe a domain-gap--diversity trade-off between synthetic and real data. These results demonstrate the value of learning synthetic stuttering data directly from real dysfluency for robust Mandarin speech processing. Code and audio examples are available at
\href{https://anonymous.4open.science/r/arr-review-artifact-B358}
{our anonymous repository}.
\end{abstract}

\section{Introduction}
Stuttering disrupts the flow of speech through phenomena such as repetitions, prolongations, and blocks. These disruptions in speech fluency pose substantial challenges to automatic speech recognition (ASR) \citep{mujtaba2024lost}. The difficulty arises not only from the irregular acoustic and temporal patterns of stuttered speech, but also from the limited availability of training data that adequately cover such variation \citep{macdonald2021disordered}. For Mandarin, AS-70 \citep{gong202470} provides an important foundation for developing stuttering-aware speech systems. Nevertheless, the availability of a real-speech corpus does not eliminate the need for broader coverage of stuttering patterns and speaking characteristics. This motivates an important practical question: how can we expand the available training data in ways that provide useful supervision beyond existing real stuttered speech?

Speech synthesis offers a promising approach to this problem, given its demonstrated utility for ASR training \citep{zheng2021using}. However, extending this approach to stuttered speech requires more than generating intelligible utterances with diverse voices: the synthesizer must reproduce irregular acoustic and temporal patterns while controlling the type and location of individual stuttering events. Strong general-purpose synthesis performance does not guarantee reliable rendering of such vocal behaviors: EmergentTTS-Eval \citep{manku2026emergenttts} identifies persistent difficulties with expressive paralinguistic cues in contemporary commercial TTS systems. Stutter-TTS \citep{zhang2022stutter} establishes the feasibility of controllable stuttering synthesis, but its reported synthesis failures and limited characterization of overall naturalness and contextual consistency leave room for further investigation. These challenges motivate a framework that jointly models local stuttering events and their broader speech context, together with a comprehensive evaluation of synthesis quality.

Beyond synthesis quality, the value of synthetic speech depends on whether it improves recognition when real stuttered speech is already available. The ASR experiments in Stutter-TTS \citep{zhang2022stutter} do not establish whether adding synthetic speech can outperform training on real stuttered speech alone. Moreover, greater similarity to the target domain may not fully characterize the usefulness of additional training samples: synthetic speech may also contribute variation that complements the real data. We therefore investigate whether closer agreement with real speech consistently translates into stronger augmentation benefits, and how this relationship changes between synthetic-only training and training that combines real and synthetic speech.

We present a Mandarin stuttering speech synthesis framework that adapts two complementary TTS paradigms using real stuttered speech. For full-utterance generation, we adapt CosyVoice2 \citep{du2024cosyvoice} through event-tag conditioning and instruction tuning. For localized reconstruction, we adapt MaskGCT \citep{wang2025maskgct} through alignment-guided masking and staged infilling training. These strategies provide different ways to combine controllable event generation with utterance-level coherence. We evaluate the generated speech through objective metrics, human perceptual assessments, and an LLM-as-a-Judge protocol. Using both adapted models, we construct synthetic AS-70 for target-domain augmentation and AISHELL1-Stutter for larger-scale augmentation, and examine their downstream utility across diverse ASR architectures. The results support the effectiveness of synthetic stuttering speech for augmentation and reveal a setting-dependent relationship between domain proximity and complementary diversity. MaskGCT exhibits a smaller domain gap and generally achieves better overall recognition performance in synthetic-only training. When synthetic speech is combined with real data, however, CosyVoice2 yields larger gains across most evaluated ASR models despite its wider domain gap, alongside greater measured diversity. These observations suggest that effective augmentation depends on both target-domain fidelity and the additional variation introduced by synthetic speech.

Our main contributions are as follows:

\begin{itemize}
    \item We present a Mandarin stuttering synthesis framework that learns from real stuttered speech through two complementary strategies: instruction- and tag-conditioned full-utterance generation, and alignment-guided reconstruction of local stuttering regions. We further conduct a comprehensive evaluation of the synthesized speech.
    \item We construct synthetic AS-70 and AISHELL1-Stutter resources, and demonstrate their augmentation utility against real-data training baselines across diverse ASR architectures.
    \item We identify a domain-gap–diversity trade-off across training settings, showing that a smaller real–synthetic domain gap does not necessarily imply greater augmentation benefits when real target-domain speech is available.
\end{itemize}

\section{Related Work}
Several public corpora have supported research on stuttered speech, including UCLASS \citep{howell2009university}, FluencyBank \citep{ratner2018fluency}, and SEP-28k \citep{lea2021sep} for English, as well as AS-70 \citep{gong202470} for Mandarin.

Synthetic stuttered speech has been explored as additional training data for stuttering event detection, including FluentNet \citep{kourkounakis2021fluentnet}, YOLO-Stutter \citep{zhou2024yolo}, and Stutter-Solver \citep{zhou2024stutter}. However, most existing approaches construct synthetic dysfluencies either through direct waveform manipulation of fluent speech \citep{kourkounakis2021fluentnet}, or by injecting predefined dysfluency patterns into text or phoneme sequences and subsequently synthesizing them with TTS systems trained primarily on fluent speech \citep{zhou2024yolo, zhou2024stutter}. Consequently, these approaches do not explicitly model stuttering patterns from naturally stuttered speech and therefore fail to capture the full acoustic and temporal variability of real stuttering events.

Among prior work, Stutter-TTS \citep{zhang2022stutter} is the most closely related work. However, its evaluation primarily focuses on improving recognition of stuttered speech while preserving performance on fluent speech. In contrast, our work investigates synthetic stuttering speech as a broader data-augmentation source, asking whether augmenting real stuttering data with synthetic counterparts can outperform training on real stuttering data alone. 

\section{Method}
\begin{figure*}[t]
  \centering
  \includegraphics[width=\linewidth]{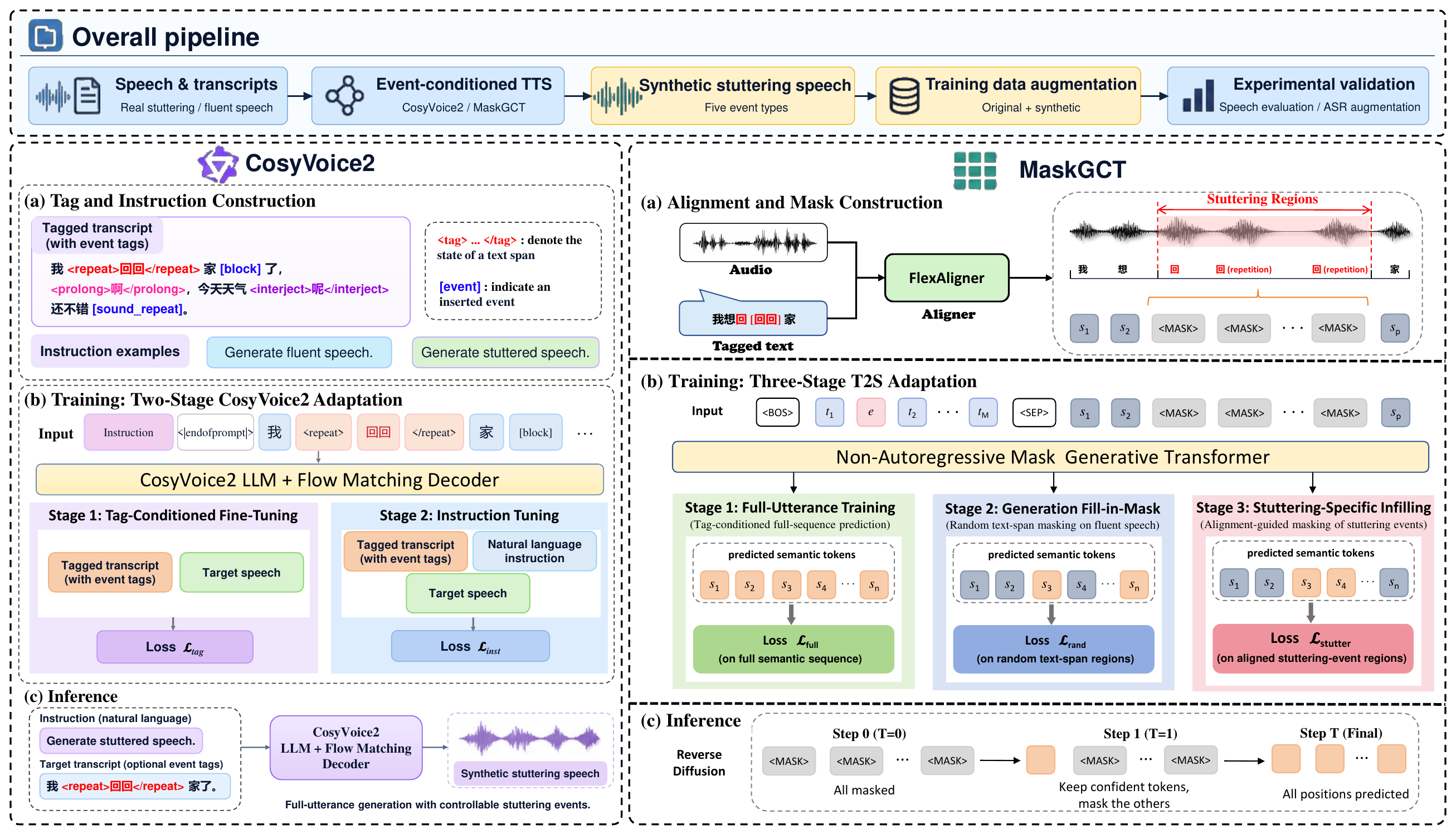}
  \caption{Overview of the proposed Mandarin stuttering speech synthesis and augmentation framework. The upper panel illustrates the overall pipeline. The lower panels show two complementary TTS adaptation strategies: CosyVoice2 employs tag-conditioned fine-tuning and instruction tuning for full-utterance generation, while MaskGCT uses alignment-guided masking and three-stage text-to-semantic (T2S) adaptation for localized stuttering generation.}
  \label{fig:overview}
\end{figure*}

Figure~\ref{fig:overview} presents the framework. We adapt CosyVoice2 and MaskGCT using naturally stuttered and fluent speech to generate Mandarin speech with five controllable stuttering event types, whose definitions are provided in Appendix~\ref{sec:stutter_events}.

\subsection{Stuttering-Aware TTS Model Adaptation}
\subsubsection{Tag-Conditioned CosyVoice2}
CosyVoice2 \citep{du2024cosyvoice} supports fine-grained instruction-based control over local speaking behaviors, providing a natural interface for introducing explicit stuttering conditions. Based on this mechanism, we extend its instruction vocabulary with five stuttering-specific tags: 
\begin{center}
\small
\begin{tabular}{c}
\texttt{<repeat>...</repeat>} \\[2pt]
\texttt{<prolong>...</prolong>} \\[2pt]
\texttt{<interject>...</interject>} \\[2pt]
\texttt{[block]} \quad
\texttt{[sound\_repeat]}
\end{tabular}
\end{center}
The paired tags enclosed by angle brackets specify the state of the corresponding textual span. In contrast, the bracketed tags denote events inserted between textual units. We first train CosyVoice2 on paired tagged transcripts and stuttered speech, enabling the model to associate each event tag with its corresponding acoustic realization.

However, stuttered speech differs from fluent speech not only in localized dysfluency events but also in broader speaking characteristics that may extend across an entire utterance. To incorporate such utterance-level information, we further perform instruction-based supervised fine-tuning using natural-language prompts that specify the desired speaking state. 

The instruction-conditioned inputs for fluent and stuttered
speech are formulated as:
\begin{equation}
\begin{aligned}
\mathbf{x}_{\mathrm{fluent}}
&= [p_{\mathrm{fluent}},
\texttt{<|endofprompt|>}, x],\\
\mathbf{x}_{\mathrm{stutter}}
&= [p_{\mathrm{stutter}},
\texttt{<|endofprompt|>}, \mathcal{T}(x,e)],
\end{aligned}
\end{equation}
where $p_{\mathrm{fluent}}$ and $p_{\mathrm{stutter}}$
denote natural-language instructions specifying fluent and
stuttered speaking states, respectively; $x$ represents the
original transcript; $e$ denotes the stuttering event annotations;
and $\mathcal{T}(x,e)$ constructs the event-tagged transcript
by inserting the corresponding stuttering tags into $x$.

Importantly, the stuttered-speech corpus also contains utterances produced by speakers who stutter that do not exhibit any explicit stuttering event. For these samples, the stuttered speaking-state instruction is provided without local event tags, allowing the model to learn broader characteristics of stuttered speech independently of explicit event conditioning. Accordingly, this design provides two complementary levels of conditioning. The natural-language instruction specifies the global speaking state, while the explicit stuttering tags determine the type and position of individual dysfluency events.

\subsubsection{Alignment-Guided MaskGCT}
\paragraph{Motivation} 
Unlike autoregressive generation, the masked-generation paradigm of MaskGCT \citep{wang2025maskgct} naturally supports localized speech infilling. This property is particularly desirable in the low-resource setting of stuttered speech, where restricting the learning objective to dysfluent regions allows the model to focus its capacity on modeling event-specific patterns without unnecessarily altering the remaining content.
\paragraph{Alignment and Mask Construction}
Unlike conventional forced alignment, our setting does not require every acoustic segment to be assigned to a textual unit. Stuttered speech may contain events that are only partially represented by the transcript, making strict one-to-one alignment unreliable. We therefore employ FlexAligner \citep{flexaligner2026}, a forced-alignment framework designed to handle imperfectly matched speech and transcripts. This ability makes it particularly suitable for stuttered speech. Detailed descriptions of FlexAligner, together with its alignment performance, are provided in Appendix~\ref{sec:adaption_details}.

Given an event-tagged transcript, we use the reliable aligned textual units immediately preceding and following each annotated stuttering event as anchors. Their temporal boundaries define an acoustic interval that is treated as the corresponding stuttering region. Importantly, this procedure does not require the anchors to coincide exactly with the nearest acoustic boundaries of the event. A slightly broader masked interval is acceptable, since the text-to-semantic model reconstructs the entire region conditioned jointly on the tagged text and the surrounding unmasked semantic context. This flexibility reduces the dependence on precise event-level timestamps while still preserving the contextual information required for localized stuttering generation.

\paragraph{Three-Stage T2S Adaptation}
Although the mask-and-predict formulation of MaskGCT naturally supports localized editing through masked token prediction, the original work reports limited robustness in speech-editing tasks. To enable more stable localized reconstruction, allowing stuttering events to be generated while maintaining consistency with the surrounding speech context, we employ a three-stage T2S adaptation strategy.
\subparagraph{Stage 1: Full-Utterance Generation.}
We first adapt the T2S model on naturally stuttered speech paired with event-tagged transcripts. The model predicts the full semantic-token sequence while conditioning on the stuttering tags, establishing tag--acoustic correspondence and adapting to broader characteristics of stuttered speech.
\subparagraph{Stage 2: General Span Infilling.}
We next perform fill-in-mask training on fluent speech by randomly selecting character- or word-level spans and masking the corresponding semantic-token regions. The model reconstructs these regions from the complete text and surrounding unmasked semantic context, thereby developing general localized reconstruction capability.
\subparagraph{Stage 3: Stuttering-Specific Infilling.}
Finally, we mask only the aligned semantic-token regions corresponding to naturally occurring stuttering events while retaining the associated event tags in the text condition. The model reconstructs the target regions from the tagged transcript and bidirectional semantic context, focusing localized generation on real stuttering patterns while preserving contextual consistency.
\subsection{Synthetic Stuttering Data Construction}
Using the two adapted TTS models described above, we construct two complementary synthetic Mandarin stuttering speech resources. The first is a resynthesized version of AS-70, where the original event-tagged transcripts are used to generate synthetic counterparts of naturally stuttered utterances. Only the AS-70 training split is resynthesized for downstream augmentation. The second, AISHELL1-Stutter, is constructed from fluent AISHELL-1 \citep{bu2017aishell} transcripts by randomly inserting stuttering events and synthesizing the resulting tagged text. Both CosyVoice2 and MaskGCT are used to generate each dataset, providing synthetic data for subsequent downstream augmentation. More details on dataset construction are provided in Appendix~\ref{sec:data_details}.

The two datasets differ mainly in how stuttering events are determined. In the synthetic AS-70 data, the event types and textual positions follow those observed in the original recordings from speakers who stutter, making the synthesized utterances more closely reflect naturally occurring stuttering patterns. In contrast, AISHELL1-Stutter introduces events at randomly selected positions in fluent transcripts, providing greater flexibility for large-scale data generation. Although its event distribution is less directly tied to natural stuttering behavior, it still serves as a useful source of additional training data for downstream augmentation.

\section{Experimental Setup}
\subsection{Evaluation Protocol}
We treat each adapted TTS pipeline as a data generator and compare the resulting synthetic datasets, not the TTS backbones. No claim is made about intrinsic synthesis capability.

We evaluate speech resynthesized from the AS-70 test set using three complementary perspectives: objective speech metrics, human subjective evaluation, and an LLM-as-a-Judge protocol following the evaluation dimensions adopted in NVV-SuperBench \citep{xue2026nvv}. The test set is excluded from TTS training, ensuring that all evaluations are conducted on unseen utterances. For objective evaluation, we report character error rate (CER) for content preservation, together with DNSMOS \citep{reddy2021dnsmos} and UTMOS(v2) \citep{baba2024t05} for speech quality. DNSMOS is further reported in terms of OVRL, SIG, BAK \citep{reddy2022dnsmos}, and P808.

For perceptual evaluation, both human listeners and Gemini 2.5 Pro \citep{comanici2025gemini} assess the speech along the same five dimensions: Naturalness, Quality, Accuracy, Perceptual Effect, and Overall Expression, using a 1--5 rating scale. The human evaluation is conducted on a small-scale subset with five listeners, each evaluating three groups of 20 utterances. Gemini 2.5 Pro serves as an audio-aware judge and evaluates speech together with the corresponding event-tagged reference transcripts.Detailed evaluation settings are provided in Appendix~\ref{sec:perceptual_evaluation}.

\subsection{ASR Fine-Tuning Setup}
We evaluate both the effectiveness of the synthesized stuttering speech itself and its utility for data augmentation across a diverse set of recognition architectures, including Mandarin-pretrained Wav2Vec 2.0 \citep{baevski2020wav2vec}, the Whisper family at four model scales \citep{radford2023robust}, Qwen3-ASR (0.6B/1.7B) \citep{shi2026qwen3}, SenseVoice \citep{an2024funaudiollm}, Paraformer \citep{gao2022paraformer}, and Fun-ASR-Nano \citep{an2025fun}. 

For both CosyVoice2- and MaskGCT-generated datasets, we consider three fine-tuning settings: Synthetic-only, trained only on resynthesized AS-70; Combined, trained on real and synthetic AS-70; and Augmented, pretrained on AISHELL1-Stutter and then fine-tuned on the combined AS-70 data. Evaluation is conducted on the AS-70 test set.

Within each model, the number of training epochs, batch size and learning rate are kept fixed across all fine-tuning settings, and the training dataset is the only variable. More details on ASR Fine-Tuning Setup are provided in Appendix~\ref{sec:detail_asr}.

\begin{table}[t]
\centering
\caption{Objective evaluation results of different speech sources and synthesis systems. Ground Truth denotes the original recordings, and MaskGCT and CosyVoice2 denote resynthesized speech.}
\label{tab:objective_eval}
\small
\setlength{\tabcolsep}{3pt}
\renewcommand{\arraystretch}{0.90}
\begin{tabular}{lcccccc}
\toprule
\multirow{2}{*}{Method}
& \multirow{2}{*}{CER}
& \multicolumn{4}{c}{DNSMOS}
& \multirow{2}{*}{UTMOS} \\
\cmidrule(lr){3-6}
& & OVRL & SIG & BAK & P808 & \\
\midrule
Ground Truth & 12.31 & 2.86 & 3.26 & 3.72 & 3.41 & 2.01 \\
MaskGCT      & 14.72 & 3.05 & 3.41 & 3.87 & 3.57 & 2.38 \\
CosyVoice2    & 10.97 & 3.10 & 3.42 & 3.98 & 3.52 & 2.26 \\
\bottomrule
\end{tabular}
\end{table}

\begin{table*}[t]
\centering
\caption{Human and Gemini 2.5 Pro perceptual evaluation of real and synthesized stuttered speech on the AS-70 test set. Both evaluations use five dimensions: Accuracy (stuttering-event realization), Perceptual Effect (perceptual effectiveness of stuttering), Naturalness, Quality, and Overall Expression (overall expressive quality). Scores range from 1 to 5, with higher values indicating better performance. Results are reported as mean ± standard deviation.}
\label{tab:mos_results}

\small
\setlength{\tabcolsep}{10pt}
\renewcommand{\arraystretch}{0.90}

\begin{tabular}{lccccc}
\toprule
Method & Accuracy & Perceptual Effect & Naturalness & Quality & Overall Expression \\
\midrule

\multicolumn{6}{c}{\textit{Human Evaluation}} \\
\midrule

Ground Truth 
& $3.87 \pm 0.35$
& $\mathbf{3.88 \pm 0.37}$
& $4.40 \pm 0.32$
& $4.17 \pm 0.33$
& $3.97 \pm 0.33$ \\

MaskGCT 
& $\mathbf{3.97 \pm 0.33}$
& $3.87 \pm 0.34$
& $4.47 \pm 0.27$
& $4.23 \pm 0.29$
& $4.03 \pm 0.30$ \\

CosyVoice2
& $3.83 \pm 0.38$
& $3.73 \pm 0.38$
& $\mathbf{4.47 \pm 0.25}$
& $\mathbf{4.67 \pm 0.25}$
& $\mathbf{4.23 \pm 0.31}$ \\

\midrule

\multicolumn{6}{c}{\textit{Gemini 2.5 Pro Evaluation}} \\
\midrule

Ground Truth 
& $3.80 \pm 0.03$
& $3.53 \pm 0.02$
& $4.02 \pm 0.03$
& $4.45 \pm 0.02$
& $3.89 \pm 0.03$ \\

MaskGCT 
& $\mathbf{3.91 \pm 0.04}$
& $\mathbf{3.55 \pm 0.04}$
& $3.88 \pm 0.03$
& $4.38 \pm 0.02$
& $3.78 \pm 0.02$ \\

CosyVoice2 
& $3.80 \pm 0.01$
& $3.51 \pm 0.03$
& $\mathbf{4.04 \pm 0.05}$
& $\mathbf{4.58 \pm 0.02}$
& $\mathbf{3.90 \pm 0.04}$ \\

\bottomrule
\end{tabular}
\end{table*}

\begin{figure*}[t]
    \centering
    \begin{subfigure}[t]{0.49\textwidth}
        \centering
        \includegraphics[width=\linewidth]{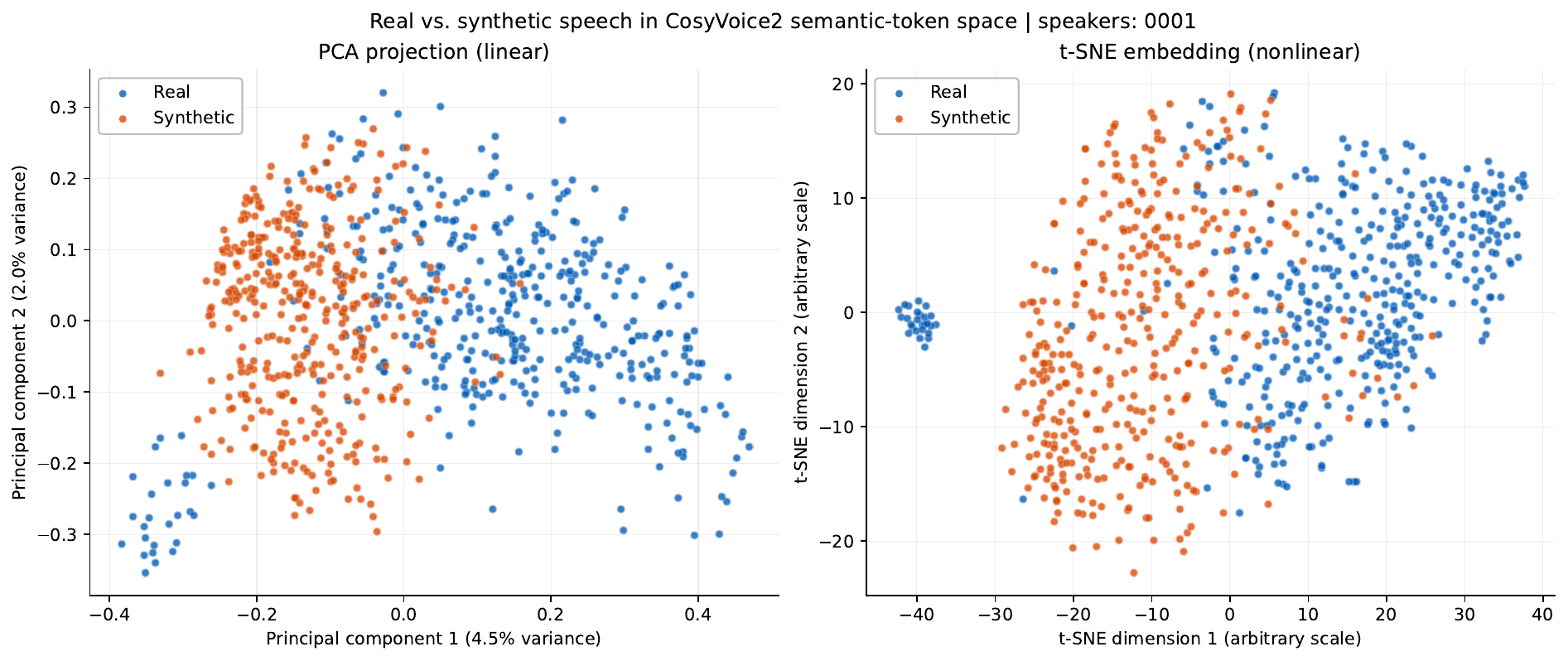}
        \caption{GT vs CosyVoice2, speaker 0001}
        \label{fig:domain_cosy_0001}
    \end{subfigure}
    \hfill
    \begin{subfigure}[t]{0.49\textwidth}
        \centering
        \includegraphics[width=\linewidth]{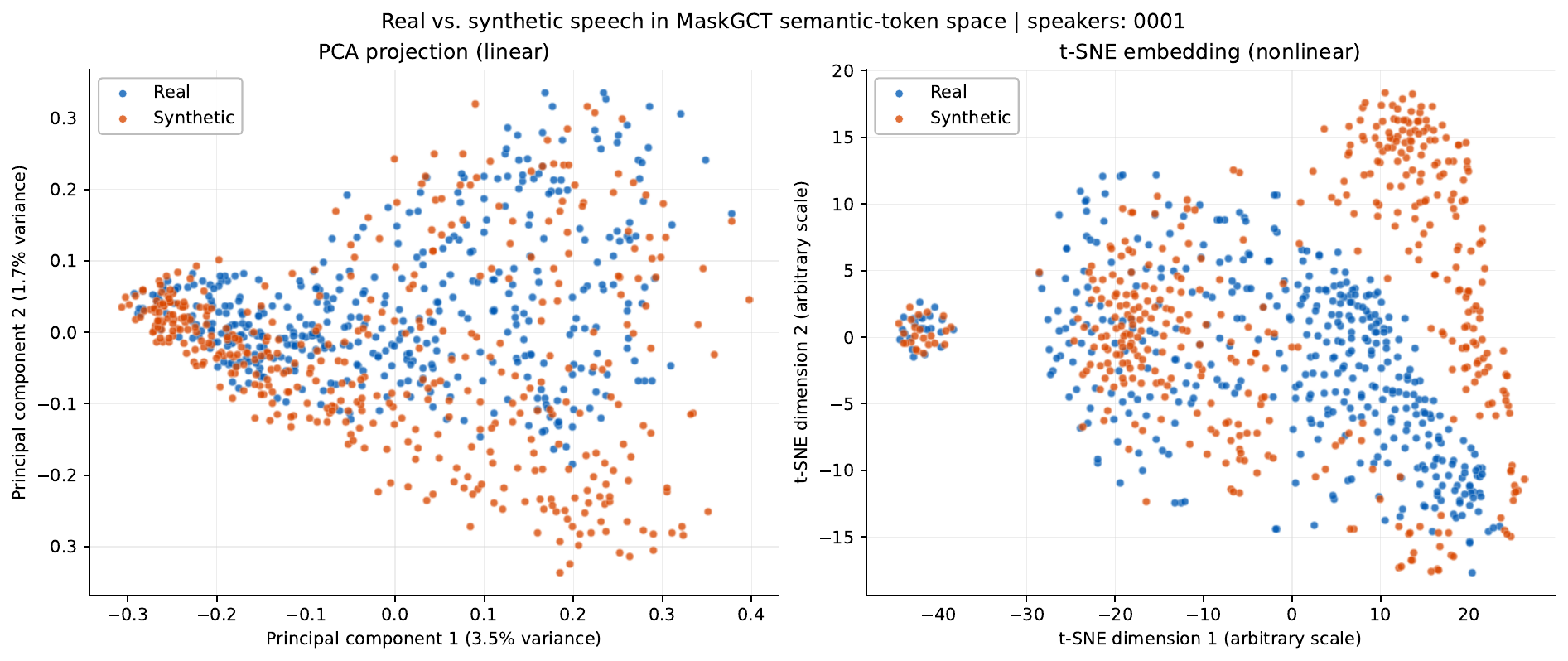}
        \caption{GT vs MaskGCT, speaker 0001}
        \label{fig:domain_mask_0001}
    \end{subfigure}

    \vspace{0.5em}

    \begin{subfigure}[t]{0.49\textwidth}
        \centering
        \includegraphics[width=\linewidth]{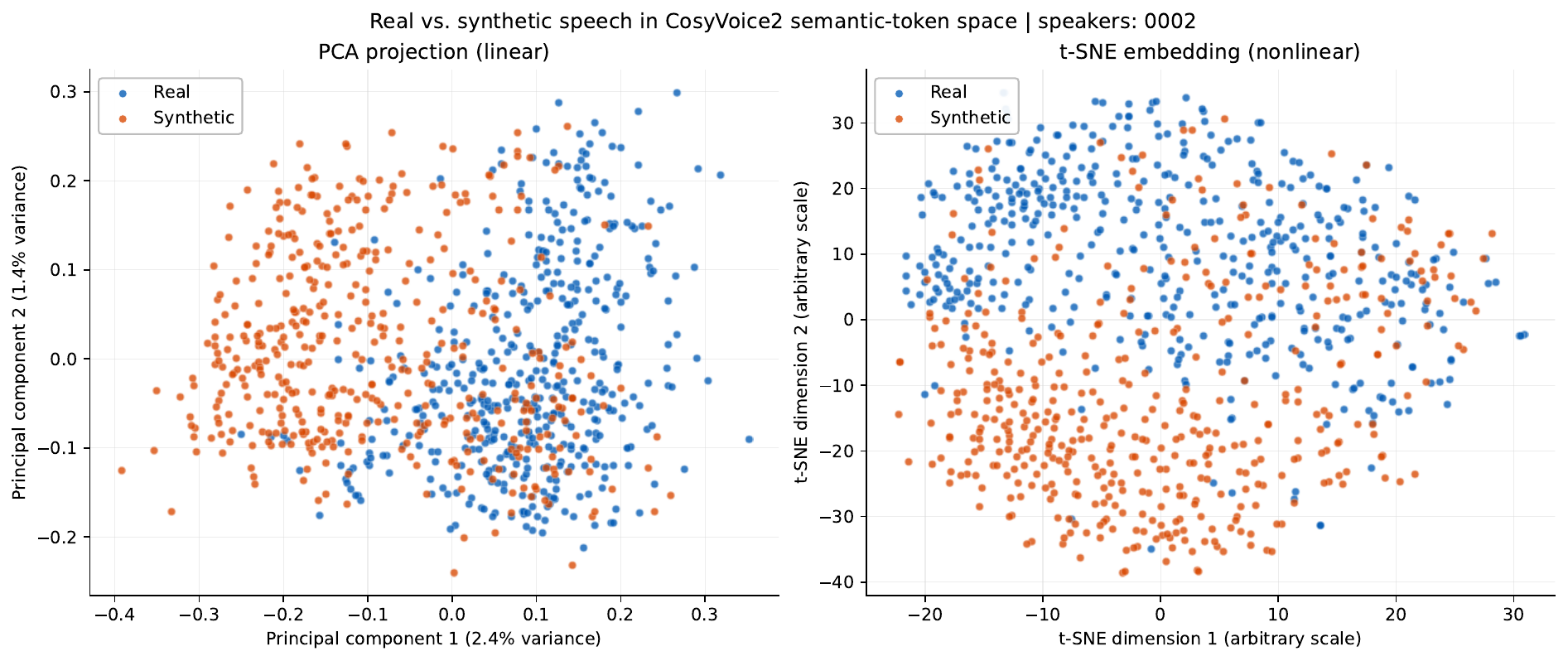}
        \caption{GT vs CosyVoice2, speaker 0002}
        \label{fig:domain_cosy_0002}
    \end{subfigure}
    \hfill
    \begin{subfigure}[t]{0.49\textwidth}
        \centering
        \includegraphics[width=\linewidth]{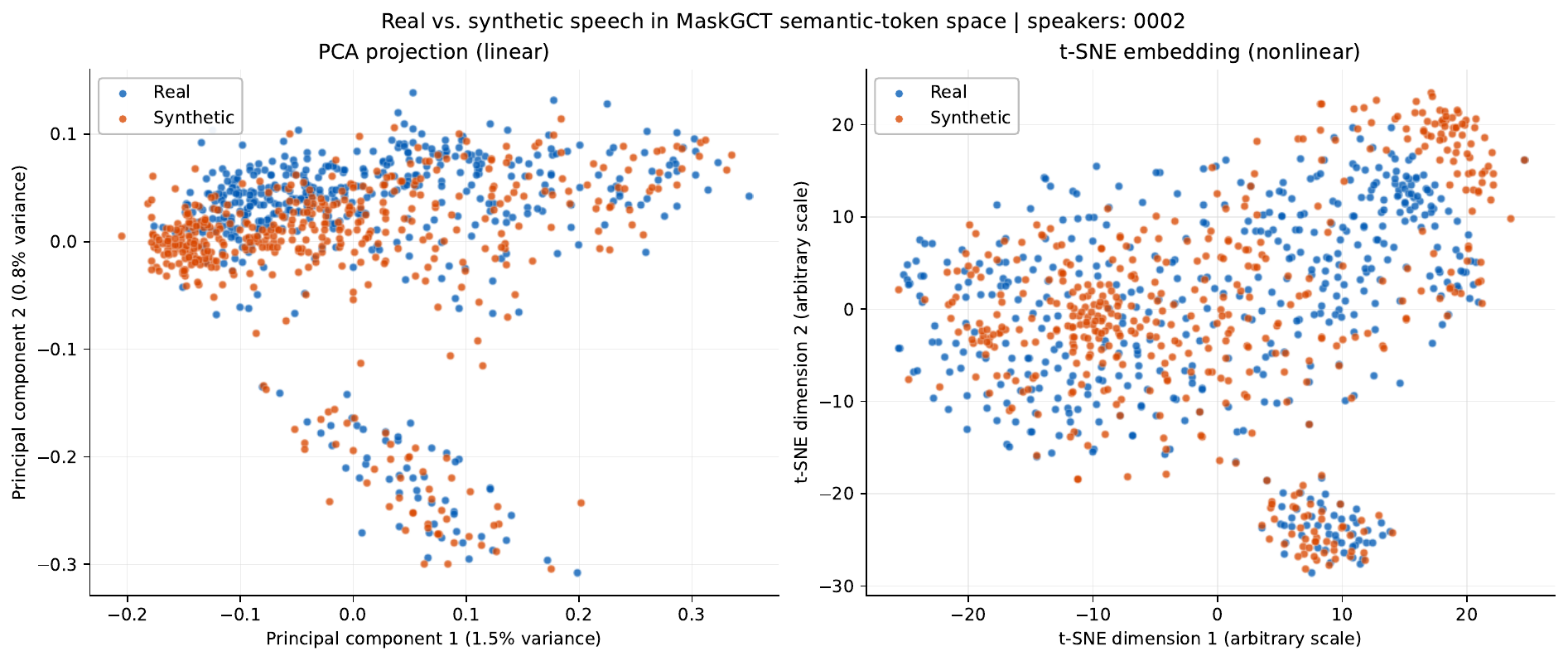}
        \caption{GT vs MaskGCT, speaker 0002}
        \label{fig:domain_mask_0002}
    \end{subfigure}

    \caption{
    PCA and t-SNE visualizations of real and synthetic speech for two
    representative speakers. Each synthesis system is analyzed in its
    corresponding semantic-token representation space.
    }
    \label{fig:domain_gap}
\end{figure*}

\section{Results and Analysis}

\subsection{Synthetic Speech Evaluation}
\subsubsection{Objective Evaluation}
Table~\ref{tab:objective_eval} reports the objective evaluation results for the synthesized speech. CosyVoice2 obtains the lowest CER, indicating strong preservation of linguistic content. Both systems outperform the ground truth across all DNSMOS dimensions, suggesting that the synthesized speech generally exhibits cleaner and more stable acoustic characteristics. A similar trend is observed for UTMOS(v2).

\subsubsection{Human Perceptual Evaluation}
Table~\ref{tab:mos_results} presents the human and the Gemini 2.5 Pro evaluation results. Both MaskGCT and CosyVoice2 achieve scores comparable to, and in several dimensions slightly higher than, Ground Truth. More importantly, the two systems exhibit complementary strengths. MaskGCT obtains the highest Accuracy score and remains competitive in Perceptual Effect, suggesting more faithful modeling and reconstruction of local stuttering events. CosyVoice2 achieves the strongest Naturalness, Quality and Overall Expression scores, indicating that full-utterance generation provides better overall speech quality and perceptual coherence.

\subsubsection{LLM-Based Perceptual Evaluation}
The Gemini 2.5 Pro results show a pattern consistent with the human judgments. MaskGCT performs best on the two stuttering-specific dimensions and CosyVoice2 achieves the strongest Naturalness, Quality, and Overall Expression. Together with the objective and human evaluations, these results suggest that both synthesized systems produce speech close to real stuttered speech, while MaskGCT is stronger in local stuttering-event realization and CosyVoice2 provides better overall perceptual quality and naturalness.

\subsection{ASR Fine-tuning Results}

\subsubsection{Representation-Space Analysis of Real--Synthetic Domain Gap}
To establish the real--synthetic domain-gap difference between the two synthesis systems, we first examine their representation distributions. Figure~\ref{fig:domain_gap} shows PCA and t-SNE projections for two representative speakers, comparing speech generated by CosyVoice2 and MaskGCT with the corresponding real AS-70 recordings.

For CosyVoice2, the real and synthetic samples exhibit relatively clear distributional separation for both speakers, particularly in the nonlinear t-SNE projections. In contrast, MaskGCT shows substantially stronger overlap between real and synthetic samples in the main distribution regions. This pattern is consistent with its localized reconstruction strategy, which modifies only the target stuttering region while preserving most of the original acoustic and contextual information.

\subsubsection{Quantitative Analysis of Domain Gap and Complementary Diversity}
\begin{figure*}[t]
    \centering

    \begin{subfigure}[t]{0.48\textwidth}
        \centering
        \includegraphics[width=\linewidth]{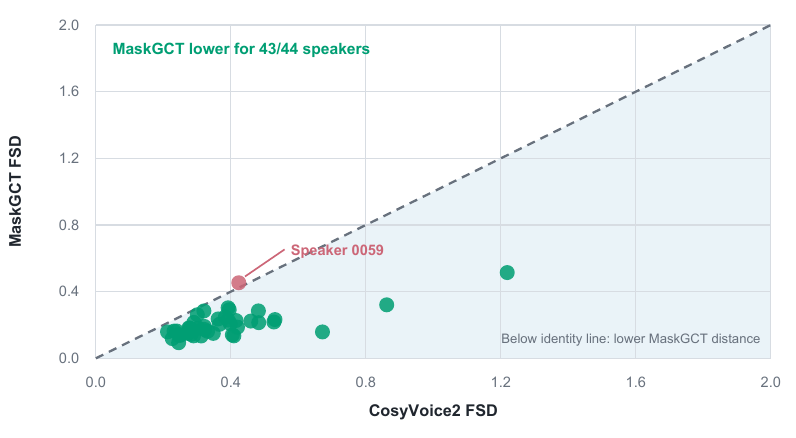}
        \caption{FSD comparison}
        \label{fig:fsd_paired}
    \end{subfigure}
    \hfill
    \begin{subfigure}[t]{0.48\textwidth}
        \centering
        \includegraphics[width=\linewidth]{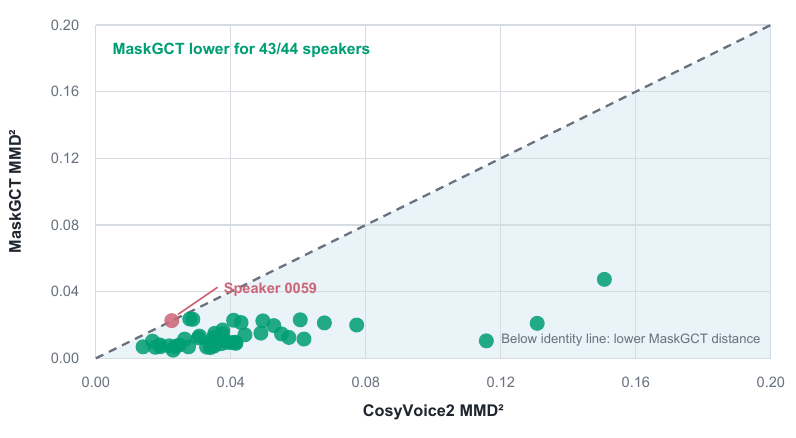}
        \caption{MMD$^2$ comparison}
        \label{fig:mmd_paired}
    \end{subfigure}

    \caption{
    Speaker-wise real--synthetic distribution distances. Points below the identity line indicate a smaller real--synthetic distribution gap for MaskGCT. MaskGCT achieves lower FSD and MMD$^2$ for 43 of the 44 speakers.
    }
    \label{fig:domain_gap_quantitative}
\end{figure*}

\begin{table}[t]
\centering
\caption{Panel (a) reports macro-averaged real--synthetic distribution distances. Panel (b) reports the relative increase in representation diversity after adding synthetic speech to the real AS-70 training data.}
\label{tab:gap_diversity_analysis}

\small
\setlength{\tabcolsep}{8pt}
\renewcommand{\arraystretch}{0.95}

\begin{tabular}{lcc}
\toprule
\multicolumn{3}{c}{\textit{(a) Real--Synthetic Domain Gap}} \\
\midrule
Method & FSD $\downarrow$ & MMD$^2$ $\downarrow$ \\
\midrule
CosyVoice2 & 0.3796 & 0.0434 \\
MaskGCT    & \textbf{0.2052} & \textbf{0.0139} \\
\bottomrule
\end{tabular}

\vspace{0.7em}

\begin{tabular}{lcc}
\toprule
\multicolumn{3}{c}{\textit{(b) Complementary Diversity Gain}} \\
\midrule
Synthetic Source
& Vendi Score $\uparrow$
& Effective Rank $\uparrow$ \\
\midrule
MaskGCT
& 3.73\%
& 14.48\% \\

CosyVoice2
& \textbf{7.37\%}
& \textbf{15.73\%} \\
\bottomrule
\end{tabular}

\end{table}

We next quantify both factors underlying the subsequent ASR analysis: target-domain proximity and complementary diversity. For domain proximity, we compute FSD \citep{kim2026fsd} and MMD$^2$ \citep{gretton2012mmd} between real and synthetic AS-70 speech using WavLM Base+ representations \citep{chen2022wavlm}. Figure~\ref{fig:domain_gap_quantitative} shows that MaskGCT yields lower distances for 43 of the 44 speakers under both metrics. Table~\ref{tab:gap_diversity_analysis} further shows its macro-averaged FSD and MMD$^2$, providing consistent quantitative evidence of a smaller domain gap.

We further measure the complementary diversity introduced when synthetic speech is added to the real AS-70 training set. Both synthesis systems increase the Vendi Score \citep{friedman2022vendi} and covariance effective rank \citep{roy2007effective}, indicating that synthetic data expand the representation diversity beyond real speech alone. CosyVoice2, however, produces consistently larger relative gains. Together, these results establish two complementary properties: MaskGCT provides stronger target-domain proximity, whereas CosyVoice2 contributes greater diversity.

\begin{table*}[t]
\centering
\caption{ASR evaluation results under different training settings.}
\label{tab:asr_results}

\setlength{\tabcolsep}{5.0pt}
\renewcommand{\arraystretch}{0.70}

\resizebox{\textwidth}{!}{
\begin{tabular}{lcccccccc}
\toprule
\multirow{2}{*}{\textbf{Model}}
& \multirow{2}{*}{\textbf{No-FT}}
& \multirow{2}{*}{\textbf{AS70-Real}}
& \multicolumn{3}{c}{\textbf{CosyVoice2}}
& \multicolumn{3}{c}{\textbf{MaskGCT}} \\
\cmidrule(lr){4-6}
\cmidrule(lr){7-9}
& &
& \textbf{AS70-Syn}
& \textbf{Combined}
& \textbf{Augmented}
& \textbf{AS70-Syn}
& \textbf{Combined}
& \textbf{Augmented} \\

\midrule
\multicolumn{9}{c}{\textit{Overall CER (\%) $\downarrow$}} \\
\midrule

Wav2Vec2-CTC
& 29.02 & 10.24 & 18.83 & 8.84 & 7.96
& 15.57 & 8.59 & \textbf{7.71} \\

Whisper-Base
& 47.86 & 20.17 & 25.47 & 19.42 & \textbf{17.26}
& 22.04 & 20.86 & 19.08 \\

Whisper-Small
& 29.73 & 8.75 & 9.23 & 7.76 & \textbf{7.45}
& 8.75 & 8.34 & 8.09 \\

Whisper-Medium
& 27.25 & 5.98 & 7.46 & 5.79 & 5.69
& 6.92 & 5.89 & \textbf{5.67} \\

Whisper-Large-v3
& 21.11 & 5.58 & 6.74 & \textbf{5.30} & \textbf{5.30}
& 6.14 & 5.44 & 5.38 \\

SenseVoice
& 13.09 & 4.05 & 4.24 & 4.02 & \textbf{3.92}
& 4.29 & 4.10 & 3.96 \\

Paraformer
& 13.25 & 5.03 & 5.12 & 4.98 & 4.91
& 5.33 & 4.88 & \textbf{4.80} \\

Fun-ASR-Nano
& 14.92 & 4.48 & 4.53 & 4.43 & \textbf{4.34}
& 4.67 & 4.49 & 4.40 \\

Qwen3-ASR-0.6B
& 12.71 & 3.98 & 4.31 & 3.86 & 3.75
& 4.09 & 3.89 & \textbf{3.73} \\

Qwen3-ASR-1.7B
& 12.31 & 3.52 & 3.87 & 3.43 & 3.28
& 3.61 & 3.48 & \textbf{3.20} \\

\midrule
\multicolumn{9}{c}{\textit{Stuttering-region Accuracy (\%) $\uparrow$}} \\
\midrule

Wav2Vec2-CTC
& 38.36 & 76.43 & 65.61 & 81.77 & 84.73
& 66.13 & 80.50 & \textbf{85.24} \\

Whisper-Base
& 62.50 & 70.43 & 70.76 & 70.14 & \textbf{71.45}
& 65.93 & 69.31 & 70.82 \\

Whisper-Small
& 68.20 & 84.03 & 83.47 & 85.58 & \textbf{86.26}
& 82.45 & 84.50 & 85.75 \\

Whisper-Medium
& 70.99 & 88.71 & 85.90 & 87.96 & \textbf{89.87}
& 86.55 & 86.74 & 89.74 \\

Whisper-Large-v3
& 69.00 & 87.46 & 88.28 & 90.63 & \textbf{90.64}
& 85.84 & 88.66 & 89.90 \\

SenseVoice
& 22.55 & 87.09 & 86.98 & 87.82 & \textbf{88.02}
& 86.60 & 86.86 & 87.01 \\

Paraformer
& 20.10 & 84.11 & 84.27 & 86.65 & 86.92
& 82.84 & 85.94 & \textbf{87.13} \\

Fun-ASR-Nano
& 22.69 & 85.59 & 85.43 & 86.98 & \textbf{87.29}
& 83.35 & 85.65 & 86.08 \\

Qwen3-ASR-0.6B
& 23.15 & 91.33 & 90.14 & 90.81 & 91.59
& 90.42 & 91.41 & \textbf{91.64} \\

Qwen3-ASR-1.7B
& 21.39 & 92.56 & 91.38 & 92.02 & 92.99
& 91.31 & 92.61 & \textbf{93.12} \\

\midrule
\multicolumn{9}{c}{\textit{Non-stuttering-region CER (\%) $\downarrow$}} \\
\midrule

Wav2Vec2-CTC
& 24.28 & 8.75 & 16.98 & 7.57 & \textbf{7.43}
& 13.40 & 7.68 & 8.16 \\

Whisper-Base
& 32.16 & 16.50 & 21.66 & 16.35 & 15.98
& 17.69 & 17.59 & \textbf{15.52} \\

Whisper-Small
& 20.25 & 7.41 & 8.28 & 6.95 & \textbf{6.82}
& 7.75 & 7.16 & 6.96 \\

Whisper-Medium
& 17.58 & \textbf{5.01} & 6.53 & 5.04 & 5.13
& 6.10 & 5.10 & 5.27 \\

Whisper-Large-v3
& 11.95 & 4.80 & 5.96 & \textbf{4.75}& 4.78
& 5.21 & 4.80 & 4.90 \\

SenseVoice
& 5.53 & 3.25 & 3.36 & 3.20 & \textbf{3.16}
& 3.34 & 3.31 & 3.23 \\

Paraformer
& 5.20 & \textbf{3.92} & 4.24 & 4.05 & 4.11
& 4.12 & \textbf{3.92} & 4.02 \\

Fun-ASR-Nano
& 4.94 & 3.41 & 3.71 & \textbf{3.40} & 3.45
& 3.53 & 3.51 & 3.57 \\

Qwen3-ASR-0.6B
& 5.16 & 3.41 & 3.66 & 3.30 & 3.29
& 3.50 & 3.32 & \textbf{3.28} \\

Qwen3-ASR-1.7B
& 4.41 & 2.97 & 3.38 & 2.87 & \textbf{2.76}
& 3.04 & 3.03 & 2.90 \\

\bottomrule
\end{tabular}
}
\end{table*}

\subsubsection{ASR Fine-tuning Results}
Table~\ref{tab:asr_results} reports ASR performance under different training settings. We evaluate overall CER, Accuracy on stuttering regions, and CER on non-stuttering regions. Event-region accuracy counts an event as correct only when the associated textual content is completely recognized without any insertion, deletion, or substitution.

Synthetic stuttering speech is consistently useful for ASR adaptation. Compared with the No-FT setting, both synthesis systems substantially improve overall CER and stuttering-region accuracy when used alone. Combining synthetic and real AS-70 data generally yields further gains, while the Augmented setting achieves the strongest overall performance for most models.

More importantly, we observe a domain-gap--diversity trade-off across augmentation settings. Under the AS70-Syn setting, MaskGCT generally achieves lower overall CER than CosyVoice2, with a more pronounced advantage in non-stuttering regions than in stuttering-region accuracy. This pattern is consistent with the preceding analysis, where MaskGCT exhibits a smaller real--synthetic domain gap. 

Interestingly, this pattern changes when synthetic speech is combined with real AS-70. Despite its larger real--synthetic domain gap, CosyVoice2-Combined outperforms MaskGCT-Combined across most ASR models. This finding is particularly noteworthy because greater target-domain similarity is generally considered desirable for synthetic training data. However, the observed results show that a smaller domain gap does not necessarily correspond to greater augmentation benefits. Instead, the stronger performance of CosyVoice2-Combined coincides with its larger complementary diversity gains. This contrast highlights the potential value of synthetic variation beyond the target-domain distribution when real speech is included in training.

The Augmented setting exhibits a related pattern. Pretraining on AISHELL1-Stutter followed by target-domain fine-tuning frequently achieves the highest stuttering-region accuracy, without a consistent deterioration in non-stuttering-region CER. Taken together, these observations support a domain-anchor interpretation: synthetic-only performance is more closely aligned with domain proximity, whereas complementary diversity becomes increasingly valuable in the presence of real target-domain data. This suggests that effective augmentation involves more than simply minimizing the real--synthetic domain gap.

\subsection{Ablation Study}
We conduct two ablation studies to determine whether synthetic stuttering speech provides additional useful information beyond simply adding more training samples. First, we duplicate the real AS-70 training data to match the data volume of the real-plus-synthetic setting. Second, to exclude the effect of AISHELL-1 itself, we pretrain the ASR models on the original fluent AISHELL-1 data before fine-tuning on real AS-70. These comparisons isolate the contribution of synthetic stuttering information from data quantity and source-corpus effects. Minor discrepancies between the ablation results and those reported in the main experiments arise from slight differences in experimental configurations. Ablation Studies on TTS Adaptation are provided in Appendix~\ref{app:tts_ablation}.
\begin{table}[t]
\centering
\caption{Ablation results on synthetic-data utility and the contribution of AISHELL1-Stutter.}
\label{tab:ablation_synthetic_utility}

\small
\setlength{\tabcolsep}{4pt}
\renewcommand{\arraystretch}{0.70}

\begin{tabular}{lccc}
\toprule
\multicolumn{4}{c}{\textit{(a) Ablation Study on Synthetic-Data Utility}} \\
\midrule
ASR Model
& Real$\times$2
& Real + CV2
& Real + MG \\
\midrule
Whisper-Small
& 8.84
& \textbf{7.73}
& 8.25 \\
Qwen3-ASR-0.6B
& 4.01
& \textbf{3.79}
& 3.94 \\
\bottomrule
\end{tabular}

\vspace{0.6em}

\begin{tabular}{lccc}
\toprule
\multicolumn{4}{c}{\textit{(b) Ablation Study on AISHELL1-Stutter Pretraining}} \\
\midrule
ASR Model
& Raw$\rightarrow$Real
& CV2$\rightarrow$Real
& MG$\rightarrow$Real \\
\midrule
Whisper-Small
& 9.06
& \textbf{7.89}
& 8.48 \\
Qwen3-ASR-0.6B
& 4.33
& 3.97
& \textbf{3.95} \\
\bottomrule
\end{tabular}

\vspace{0.3em}

\begin{minipage}{0.98\linewidth}
\footnotesize
\textit{Note:} CV2 and MG denote CosyVoice2 and MaskGCT; Raw denotes AISHELL-1.
\end{minipage}

\end{table}

\section{Conclusion}
We presented a framework for synthesizing Mandarin stuttered speech and demonstrated its utility for ASR augmentation. Our results suggest that target-domain proximity alone does not determine augmentation benefits: when real speech is available, complementary synthetic variation can provide additional gains. These findings highlight the value of considering both fidelity and diversity when constructing synthetic training data.

\section*{Limitations}
The two TTS pipelines differ substantially in their architectures and generation mechanisms. They are employed as complementary approaches to constructing synthetic stuttering speech for ASR augmentation and producing datasets with distinct distributional characteristics and degrees of diversity. Our objective is therefore to investigate the properties and downstream utility of these synthetic datasets, rather than to establish a controlled comparison of the intrinsic synthesis capabilities. Moreover, the scale of our human perceptual evaluation remains limited. Furthermore, the current framework is instantiated with only two representative TTS backbones. Broader evaluation across additional synthesis architectures is needed to more fully assess the generalizability of the framework. Finally, the domain-gap--diversity trade-off reported in this work is a counterintuitive empirical observation rather than a causally established mechanism. A rigorous investigation of the mechanisms underlying this phenomenon remains an important direction for future research.

\section*{Ethical Considerations}
This work uses existing speech corpora and follows their respective licenses and usage conditions. The study focuses on synthesizing stuttering speech for research on speech recognition, with the goal of improving speech technologies for speakers who stutter. Since speech synthesis may potentially reproduce speaker-specific characteristics, the generated speech should not be used for impersonation or other identity-related applications without appropriate consent. In addition, stuttering is a speech disorder with substantial individual variability; the synthesized speech is intended for data augmentation and research purposes and should not be interpreted as representing the full diversity of real-world stuttering or as a clinical characterization of individuals who stutter.


\bibliography{custom}

\clearpage
\appendix

\section{Method Details}
\label{sec:method}

\subsection{Stuttering Event Definitions}
\label{sec:stutter_events}
We follow the stuttering event taxonomy adopted in AS-70 \citep{gong202470} and use five event categories throughout this work. These tags are used consistently for model conditioning, synthetic data construction, and downstream event annotation. Their definitions are summarized as follows:
\begin{itemize}
\item \textbf{[...]: Word/phrase repetition.} Repetition of an entire character or phrase.
\item \textbf{/b: Block.} Gasps for air or stuttered pauses.
\item \textbf{/p: Prolongation.} Abnormal elongation of a phoneme.
\item \textbf{/r: Sound repetition.} Repeated phonemes that do not constitute an entire character.
\item \textbf{/i: Interjection.} Filler sounds or characters introduced by stuttering. Naturally occurring interjections that do not disrupt speech fluency are excluded.
\end{itemize}

\subsection{Model Adaptation Details}
\label{sec:adaption_details}
\subsubsection{Adapted CosyVoice2}
For the instruction-based adaptation of CosyVoice2, we jointly use AS-70 and AISHELL-3 \citep{shi2020aishell} to provide supervision for stuttered and fluent speaking states, respectively. AS-70 provides naturally stuttered speech paired with event-tagged transcripts, while AISHELL-3 is used as fluent-speech supervision. During training, stuttered utterances are paired with the stuttered speaking-state instruction, whereas AISHELL-3 utterances are paired with the fluent speaking-state instruction. This joint training enables the model to learn the distinction between fluent and stuttered speaking styles while retaining explicit control over local stuttering events through event tags.

We primarily fine-tune the text-speech LM of CosyVoice2. In addition, because naturally stuttered speech is likely underrepresented in the original CosyVoice2 training data, we further adapt the downstream chunk-aware causal flow-matching model using AS-70 stuttered speech. This additional acoustic-stage adaptation is intended to better preserve dysfluency-related acoustic and temporal characteristics during waveform generation.

\subsubsection{Adapted MaskGCT}
\paragraph{FlexAligner Details}
FlexAligner is a wav2vec 2.0-based \citep{baevski2020wav2vec} forced-alignment framework designed for speech–text pairs with imperfect correspondence. It adopts a two-stage alignment strategy. The first stage performs CTC-based chunking to identify reliable matched regions while allowing unmatched acoustic segments to remain unaligned. The second stage performs local alignment within these reliable regions to obtain fine-grained temporal boundaries, followed by boundary refinement and calibration. This design enables the aligner to tolerate insertions, omissions, and other local speech–text mismatches that are common in spontaneous or dysfluent speech. The GitHub repository of this model has been made open-source \citep{flexaligner2026}.

According to information provided by the FlexAligner authors, the model is trained using LibriSpeech clean-100 \citep{panayotov2015librispeech} together with speech from the SCOTUS corpus \citep{yuan2008speaker}. Its alignment performance is evaluated on TIMIT \citep{garofolo1993darpa} and Buckeye \citep{pitt2005buckeye}. On TIMIT, evaluation focuses on local boundary accuracy, including average boundary error and the proportion of predicted boundaries falling within several temporal tolerances. On Buckeye, word-level alignment is evaluated by aligning transcript segments with the corresponding speech recordings. These evaluations indicate that FlexAligner provides accurate local boundary estimates while remaining robust to imperfect speech–text correspondence. We contacted the authors for additional technical details; however, the corresponding paper is not yet publicly available. We therefore refrain from reporting unpublished implementation details or quantitative results beyond the information that has been publicly released or explicitly shared for reference.

In our setting, FlexAligner is not used to obtain exact boundaries for the stuttering event itself. Instead, we identify reliable aligned textual units immediately before and after each annotated event and use their temporal boundaries as anchors for constructing the masked semantic-token region. This is particularly suitable for stuttered speech, where repetitions, blocks, prolongations, and other dysfluencies may not exhibit a clean one-to-one correspondence with the transcript. Because MaskGCT reconstructs the entire masked region conditioned on both the tagged text and surrounding semantic context, a slightly broader interval does not require precise event-level boundary estimation.
\paragraph{Inference Details}
Due to space constraints, we omit the detailed inference procedure of the adapted MaskGCT from the main text. As a non-autoregressive model, MaskGCT requires the target semantic-token sequence length to be specified before generation. We therefore adopt different inference strategies for the two synthetic datasets.

For the resynthesized AS-70 data, the original utterances already contain naturally occurring stuttering events. After alignment, we directly mask the semantic-token regions corresponding to the target events and regenerate these regions while preserving the surrounding context. Therefore, no additional duration estimation is required.

For AISHELL1-Stutter, the source AISHELL-1 utterances are fluent and contain no existing stuttering regions. We first estimate the duration of each inserted stuttering event before constructing the masked sequence. Specifically, we obtain empirical duration ranges for each event type from naturally stuttered speech and randomly sample a duration within the corresponding range. The sampled duration is further scaled according to the speaking rate of the target utterance, and a mask sequence of the resulting length is inserted at the designated event position for localized generation.

This rule-based duration estimation provides a simple way to construct stuttering regions without requiring additional temporal annotations, but it cannot fully capture the context-dependent duration variability of natural stuttering. A dedicated duration predictor trained on stuttered speech would provide a more principled solution and is left for future work.
\subsection{Data Construction Details}
\label{sec:data_details}
Table~\ref{tab:data_construction} summarizes the datasets used in our synthesis and downstream experiments. Synthetic AS-70 is constructed only from the original AS-70 training split, while the AS-70 test split is kept unchanged and used exclusively for evaluation. CosyVoice2 successfully generates 23,232 of the 24,395 training utterances, with the remaining samples discarded due to synthesis failures, whereas MaskGCT generates the full training split. AISHELL1-Stutter is constructed from the first 50,000 utterances of AISHELL-1, and separate versions are synthesized using CosyVoice2 and MaskGCT. The synthetic AS-70 data are used for both synthesis evaluation and downstream ASR augmentation, while AISHELL1-Stutter is used for large-scale augmentation prior to AS-70 fine-tuning.
\begin{table*}[t]
\centering
\small
\caption{Summary of the datasets used for synthetic data construction and downstream evaluation.}
\label{tab:data_construction}
\begin{tabular}{llrl}
\toprule
Dataset & Synthesis Model & \# Utterances & Usage \\
\midrule
AS-70 (train) & -- & 24,395 & TTS adaptation / Downstream training \\
AS-70 (test)  & -- & 8,514  & Evaluation only \\
\midrule
Synthetic AS-70 & CosyVoice2 & 23,232 & Downstream augmentation \\
Synthetic AS-70 & MaskGCT    & 24,395 & Downstream augmentation \\
\midrule
AISHELL1-Stutter & CosyVoice2 & 50,000 & Large-scale ASR pretraining \\
AISHELL1-Stutter & MaskGCT    & 50,000 & Large-scale ASR pretraining \\
\bottomrule
\end{tabular}
\end{table*}

\section{Experimental Setup Details}
\subsection{Details of TTS Adaptation}
\subsubsection{CosyVoice2 Fine-Tuning Details}
We adopt CosyVoice2-0.5B as the backbone and perform two-stage fine-tuning of its language model (LLM), while keeping the flow-matching decoder and vocoder frozen. Both stages use the Adam optimizer with learning rates of $1 \times 10^{-5}$ and $5 \times 10^{-6}$, respectively. We adopt a constant learning-rate scheduler configured with 2,500 warmup steps. Training uses dynamic batching with a maximum batch-frame budget of 2,000. The audio sampling rate is 24~kHz, and the acoustic features comprise 80-dimensional Mel spectrograms extracted with a hop size of 480 samples. In practice, we observe that only 2--3 epochs of fine-tuning are sufficient for each stage to achieve satisfactory synthesis performance.

\subsubsection{MaskGCT Fine-Tuning Details}
We adopt MaskGCT as the backbone and perform three-stage adaptation of its text-to-semantic (T2S) model. All three stages use the AdamW optimizer with a learning rate of $1 \times 10^{-5}$ and a batch size of 8. The input audio is sampled at 16~kHz, and the semantic tokens are represented at a rate of 50 frames per second. The synthesized waveform is reconstructed at 24~kHz with a hop size of 480 samples. All experiments were conducted on NVIDIA A800 GPUs with 80GB of VRAM.

\subsection{Details of Human Perceptual Evaluation}
The human perceptual evaluation uses the same five rating dimensions as the LLM-as-a-Judge evaluation: Naturalness, Quality, Stuttering Accuracy, Stuttering Perceptual Effect, and Overall Expression. All dimensions are rated on a 1--5 scale. Their detailed definitions and scoring criteria are identical to those specified in the LLM-as-a-Judge prompt provided below.

Five listeners participate in the evaluation. Each listener evaluates the same set of 60 utterances, consisting of 20 matched samples from each of three speech sources: the original AS-70 recordings (Ground Truth), MaskGCT, and CosyVoice2. The 20 linguistic contents and corresponding stuttering-event annotations are shared across the three sources; that is, for each selected reference utterance, the evaluation set contains its original recording together with the corresponding MaskGCT- and CosyVoice2-generated versions. This paired design ensures that differences across systems are evaluated under identical textual content and target stuttering conditions.

For each listener, all 60 utterances are pooled and randomly shuffled before evaluation. System identities are hidden, and listeners are not informed whether an utterance originates from Ground Truth, MaskGCT, or CosyVoice2. After all ratings are collected, we compute the mean and standard deviation for each of the five dimensions separately for the three speech sources.
\subsection{Details of LLM-as-Judge Perceptual Evaluation}
\label{sec:perceptual_evaluation}

The following prompt is used for all Gemini 2.5 Pro evaluations.
The original prompt is written in Chinese because the evaluated speech
and reference transcripts are in Mandarin. For readability, we provide
an English translation below.

\begin{promptbox}{LLM-as-Judge Prompt (Gemini 2.5 Pro)}

You are a professional evaluator of synthesized Mandarin stuttered speech.
You will be given a synthesized Mandarin utterance together with its
corresponding reference transcript, which contains the expected stuttering
events and their annotations. Your task is to listen to the complete audio
and evaluate both the overall speech quality and the correctness and
perceptual realism of the target stuttering events.

\medskip
\textbf{1. Stuttering Event Definitions}

\begin{itemize}[leftmargin=*,nosep]

    \item \textbf{Word/Phrase Repetition: \texttt{[...]}} \\
    Square brackets indicate repetition of a complete Chinese character,
    word, or phrase. The synthesized speech should contain a clearly audible repetition
    of the corresponding linguistic content.

    \item \textbf{Block: \texttt{/b}} \\
    \texttt{/b} denotes a block, where the speaker experiences an
    involuntary difficulty in initiating or continuing phonation,
    resulting in a temporary interruption of speech flow.

    \item \textbf{Prolongation: \texttt{/p}} \\
    \texttt{/p} denotes prolongation, where a speech sound or phonetic
    component is involuntarily sustained for noticeably longer than in
    normal pronunciation.

    \item \textbf{Sound Repetition: \texttt{/r}} \\
    \texttt{/r} denotes sound repetition, where a sub-character speech
    unit is involuntarily repeated, such as a phoneme, onset, rhyme
    component, or another speech unit that does not constitute a complete
    Chinese character.

    \item \textbf{Interjection: \texttt{/i}} \\
    \texttt{/i} denotes a stuttering-related interjection, referring to
    filler or hesitation sounds inserted outside the normal linguistic
    content, such as ``um'', ``ah'', ``uh'', or similar hesitation sounds.

\end{itemize}

\medskip
\textbf{2. Evaluation Dimensions}

Please independently rate the audio along the following five dimensions.
Each dimension should be assigned an integer score from 1 to 5.

\begin{itemize}[leftmargin=*,nosep]

    \item \textbf{A. Naturalness} \\
    Evaluate whether the complete utterance sounds natural and resembles
    real human speech. Stuttering is intentionally generated in this task
    and should therefore not reduce the naturalness score merely because
    dysfluency is present.

    \item \textbf{B. Quality} \\
    Evaluate the acoustic and technical quality of the synthesized speech,
    with emphasis on clarity, background noise, distortion, and other
    noticeable acoustic artifacts.

    \item \textbf{C. Stuttering Accuracy} \\
    Evaluate whether the target stuttering event specified in the reference
    transcript is correctly realized in the audio. Consider whether the
    event is actually present, whether the stuttering type is correct,
    whether it occurs at approximately the intended position, and whether
    any obvious unintended stuttering events are introduced.

    \item \textbf{D. Stuttering Perceptual Effect} \\
    Evaluate whether the target stuttering event sounds perceptually
    realistic and clearly identifiable. Consider whether the event resembles
    natural stuttering, whether it contains obvious mechanical repetition or
    splicing artifacts, whether transitions before and after the event are
    natural, and whether the event integrates smoothly into the utterance.

    \item \textbf{E. Overall Expression} \\
    Evaluate the overall naturalness, consistency, and coordination of the
    utterance in terms of speaking rate, rhythm, prosody, expression, and
    integration of the stuttering event.

\end{itemize}

\medskip
\textbf{3. Scoring Criteria}

All five dimensions use the following general 1--5 scale:

\begin{itemize}[leftmargin=*,nosep]
    \item \textbf{5 --- Excellent:} Very strong performance with high
    naturalness, correctness, or realism, and no obvious perceptual problems.
    
    \item \textbf{4 --- Good:} Generally strong performance with only minor
    issues that do not substantially affect the overall impression.
    
    \item \textbf{3 --- Fair:} Acceptable overall, but with clearly
    perceptible problems.
    
    \item \textbf{2 --- Poor:} Serious problems are present and noticeably
    degrade performance on the corresponding dimension.
    
    \item \textbf{1 --- Very Poor:} Severe errors, strong unnaturalness,
    serious distortion, or complete absence of the target phenomenon.
\end{itemize}

\medskip
\textbf{Please strictly follow the rules below:}

\begin{enumerate}[leftmargin=*,nosep]
    \item Listen to the complete audio before assigning any score.
    
    \item The reference transcript and stuttering annotations are provided
    only to specify the intended linguistic content, stuttering type, and
    approximate event position.
    
    \item Do not assume that a stuttering event has been successfully
    generated merely because it is marked in the transcript. The judgment
    must be based on the actual audio.
    
    \item Score all five dimensions independently.
    
    \item Do not infer acoustic realization solely from the textual
    annotations. Final judgments must rely on audible evidence in the audio.
    
    \item Use the full 1--5 scoring range when appropriate, and avoid
    systematically assigning similar or overly high scores across samples.
\end{enumerate}

\medskip
\textbf{4. Output Format}

Return only one valid JSON object. Do not output Markdown, explanations,
or any text outside the JSON object.

Use exactly the following format:

\begin{lstlisting}[style=jsonstyle]
{
  "naturalness": {
    "score": 1,
    "reason": "Briefly describe the naturalness of the utterance."
  },
  "quality": {
    "score": 1,
    "reason": "Briefly describe the acoustic quality of the audio."
  },
  "stuttering_accuracy": {
    "score": 1,
    "reason": "Briefly describe whether the target stuttering type, content, and approximate position are correctly realized."
  },
  "stuttering_perceptual_effect": {
    "score": 1,
    "reason": "Briefly describe whether the target stuttering event is realistic, clearly perceptible, and naturally integrated with the surrounding speech."
  },
  "overall_expression": {
    "score": 1,
    "reason": "Briefly describe the overall speaking rate, rhythm, prosody, expression, and integration of the stuttering event."
  }
}
\end{lstlisting}

All \texttt{score} values must be integers between 1 and 5.

\end{promptbox}

To reduce potential evaluation bias and improve stability, we adopt two additional controls. First, all candidate systems are anonymized and randomized: audio samples are presented under anonymous identifiers (A/B/C), and their presentation order is independently randomized to prevent system identity or ordering from influencing the judgments. Second, we perform repeated large-scale evaluation: for each synthesis system and the corresponding ground-truth recordings, we sample 5,000 utterances with matched transcripts and stuttering-event annotations. These samples are divided into five non-overlapping subsets of 1,000 utterances and evaluated separately. We report the mean and standard deviation calculated across the five subset-level results, providing a more stable estimate than evaluation on a single sample set.

\begin{table*}[t]
\centering
\caption{LLM-as-a-Judge ablation results for the adaptation strategies of CosyVoice2 and MaskGCT. Both experiments are evaluated using Gemini 2.5 Pro across five perceptual dimensions on a 1--5 scale ($\uparrow$). Results are reported as mean $\pm$ standard deviation.
}
\label{tab:tts_adaptation_ablation}

\small
\setlength{\tabcolsep}{8pt}
\renewcommand{\arraystretch}{1.05}

\begin{tabular}{lccccc}
\toprule
Method
& Accuracy
& Perceptual Effect
& Naturalness
& Quality
& Overall Expression \\
\midrule

\multicolumn{6}{c}{
\textit{(a) CosyVoice2: Instruction Conditioning}
} \\
\midrule
Single instruction
& 3.79 $\pm$ 0.03
& 3.45 $\pm$ 0.02
& 4.01 $\pm$ 0.03
& 4.59 $\pm$ 0.02
& 3.79 $\pm$ 0.02 \\

Dual instructions
& 3.80 $\pm$ 0.01
& \textbf{3.51 $\pm$ 0.03}
& 4.04 $\pm$ 0.05
& 4.58 $\pm$ 0.02
& \textbf{3.90 $\pm$ 0.04} \\

\midrule
\multicolumn{6}{c}{
\textit{(b) MaskGCT: Masking Strategy}
} \\
\midrule

Random masking
& 3.84 $\pm$ 0.03
& 3.43 $\pm$ 0.04
& 3.86 $\pm$ 0.05
& 4.32 $\pm$ 0.01
& 3.81 $\pm$ 0.04 \\

Stuttering-region masking
& \textbf{3.91 $\pm$ 0.04}
& \textbf{3.55 $\pm$ 0.04}
& 3.88 $\pm$ 0.03
& 4.38 $\pm$ 0.02
& 3.78 $\pm$ 0.02 \\

\bottomrule
\end{tabular}
\end{table*}

\begin{table*}[t]
\centering
\caption{
Detailed representation-diversity analysis after augmenting the real
AS-70 training data with synthetic speech. Absolute gain is computed
relative to the real-only set, and relative gain is the speaker-wise
percentage increase. Higher values indicate greater representation
diversity.
}
\label{tab:appendix_diversity_detail}

\small
\setlength{\tabcolsep}{7pt}
\renewcommand{\arraystretch}{0.95}

\begin{tabular}{lcccccc}
\toprule
\multirow{2}{*}{Training Data}
& \multicolumn{3}{c}{Vendi Score}
& \multicolumn{3}{c}{Covariance Effective Rank} \\
\cmidrule(lr){2-4}
\cmidrule(lr){5-7}
& Value
& Abs. Gain
& Rel. Gain
& Value
& Abs. Gain
& Rel. Gain \\
\midrule

Real only
& 4.0742
& --
& --
& 38.4616
& --
& -- \\

Real + MaskGCT
& 4.2085
& +0.1343
& +3.73\%
& 43.8603
& +5.3987
& +14.48\% \\

Real + CosyVoice2
& \textbf{4.3789}
& \textbf{+0.3046}
& \textbf{+7.37\%}
& \textbf{44.1198}
& \textbf{+5.6582}
& \textbf{+15.73\%} \\

\bottomrule
\end{tabular}
\end{table*}

\begin{figure*}[t]
    \centering
    \begin{subfigure}[t]{0.48\textwidth}
        \centering
        \includegraphics[width=\linewidth]
        {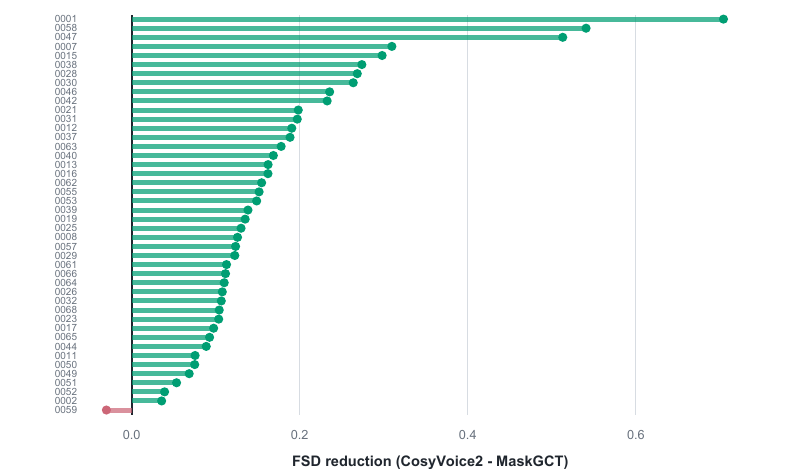}
        \caption{Speaker-wise FSD reduction}
        \label{fig:appendix_fsd_reduction}
    \end{subfigure}
    \hfill
    \begin{subfigure}[t]{0.48\textwidth}
        \centering
        \includegraphics[width=\linewidth]
        {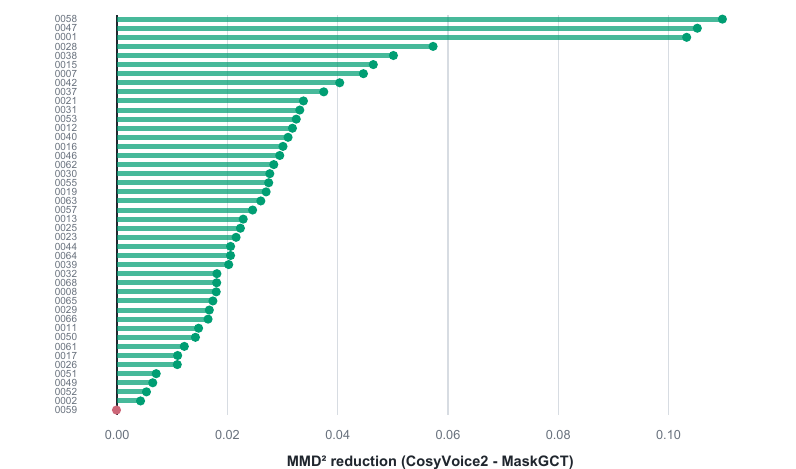}
        \caption{Speaker-wise MMD$^2$ reduction}
        \label{fig:appendix_mmd_reduction}
    \end{subfigure}

    \caption{
    Speaker-wise reductions in real--synthetic
    distribution distances across 44 AS-70 speakers,
    defined as the CosyVoice2 distance minus the
    MaskGCT distance. Positive values indicate a
    smaller domain gap for MaskGCT. Both metrics
    favor MaskGCT for 43 of the 44 speakers.
    }
    \label{fig:appendix_domain_reduction}
\end{figure*}

\begin{table*}[t]
\centering
\caption{
Strict recognition accuracy (\%) for five stuttering-event types. Each cell reports \textbf{AS70-Real / MaskGCT-Combined}, and the better result is highlighted in bold. Numbers in parentheses indicate the number of events in the test set. An event is counted as correct only when its associated textual content is recognized exactly, without insertion, deletion, or substitution within or adjacent to the event span. For \texttt{/i}, omission of the interjection itself is counted as correct.
}
\small
\setlength{\tabcolsep}{5pt}

\begin{tabular}{lrrrrrr}
\toprule
\textbf{ASR Model}
& \textbf{/b} (695)
& \textbf{/p} (1980)
& \textbf{/r} (1216)
& \textbf{/i} (1672)
& \textbf{[$\ldots$]} (4921)
& \textbf{Overall} \\
\midrule

Fun-ASR-Nano
& 90.79 / \textbf{91.08}
& \textbf{88.48} / 88.23
& 85.69 / \textbf{86.27}
& \textbf{81.88} / 76.67
& 84.92 / \textbf{86.75}
& 85.59 / \textbf{85.65} \\

Paraformer
& 89.50 / \textbf{89.78}
& 86.62 / \textbf{87.58}
& 83.88 / \textbf{84.79}
& 84.15 / \textbf{88.58}
& 82.38 / \textbf{84.13}
& 84.11 / \textbf{85.94} \\

SenseVoice
& 90.36 / \textbf{90.79}
& \textbf{89.90} / 88.79
& \textbf{85.12} / 84.79
& 88.70 / \textbf{88.94}
& \textbf{85.45} / 85.33
& \textbf{87.09} / 86.86 \\

\midrule

Wav2Vec2-CTC
& 79.28 / \textbf{80.58}
& 79.34 / \textbf{83.79}
& 75.33 / \textbf{77.96}
& 86.30 / \textbf{91.39}
& 71.77 / \textbf{76.10}
& 76.43 / \textbf{80.50} \\

\midrule

Qwen3-ASR-0.6B
& \textbf{92.52} / 92.37
& 91.36 / \textbf{91.77}
& \textbf{90.13} / 89.97
& 90.97 / \textbf{92.70}
& \textbf{91.57} / 91.04
& 91.33 / \textbf{91.41} \\

Qwen3-ASR-1.7B
& \textbf{92.81} / 91.80
& \textbf{92.53} / 92.47
& \textbf{91.86} / 91.61
& 92.11 / \textbf{93.54}
& \textbf{92.87} / 92.70
& 92.56 / \textbf{92.61} \\

\midrule

Whisper-Base
& \textbf{73.09} / 72.81
& \textbf{72.88} / 72.27
& \textbf{63.32} / 62.01
& \textbf{87.20} / 84.09
& \textbf{65.13} / 64.40
& \textbf{70.43} / 69.31 \\

Whisper-Small
& \textbf{84.03} / 83.02
& 84.34 / \textbf{86.01}
& \textbf{80.35} / 80.26
& 90.13 / \textbf{90.25}
& 82.75 / \textbf{83.19}
& 84.03 / \textbf{84.50} \\

Whisper-Medium
& \textbf{89.35} / 88.63
& \textbf{90.51} / 88.59
& \textbf{87.58} / 85.03
& \textbf{87.44} / 84.99
& \textbf{88.60} / 86.75
& \textbf{88.71} / 86.74 \\

Whisper-Large-v3
& \textbf{89.64} / 88.49
& 88.69 / \textbf{89.65}
& 85.69 / \textbf{87.34}
& \textbf{89.65} / 88.16
& 86.34 / \textbf{88.78}
& 87.46 / \textbf{88.66} \\

\bottomrule
\end{tabular}

\label{tab:appendix_five_event_accuracy}
\end{table*}

\subsection{Details of ASR Fine-Tuning Setup}
\label{sec:detail_asr}
We fine-tune four families of ASR models using their respective training frameworks. For each ASR model, we maintain the same training configuration across data conditions to ensure consistent comparisons. The model-specific configurations are detailed below.

We fine-tune the Mandarin-pretrained Wav2Vec2-CTC model using SpeechBrain with a maximum of 15 training epochs and a batch size of 8. Audio is sampled at 16~kHz. The Wav2Vec2 encoder remains trainable and is optimized using Adam with a learning rate of $5 \times 10^{-5}$. The additional recognition layers are optimized using Adadelta with a learning rate of $0.5$. Both optimizers employ NewBob learning-rate scheduling. Training uses FP32 precision, and inference
adopts CTC beam search with a beam size of 100.

We fine-tune four Whisper variants using the Hugging Face Seq2SeqTrainer with Chinese transcription as the target task. Training is configured for up to 10 epochs with a learning rate of $1 \times 10^{-5}$, a per-device batch size of 8. We use 500 warmup steps, BF16 mixed precision, and gradient checkpointing.  Early stopping is applied with a patience of ten validation evaluations.

We fine-tune SenseVoice, Paraformer, and Fun-ASR-Nano using the FunASR training framework with identical training configurations. All three models are trained for up to 20 epochs with a learning rate of $2 \times 10^{-4}$. We employ token-based dynamic batching with a maximum batch budget of 6,000 tokens.

We fine-tune Qwen3-ASR using the Hugging Face Trainer with a learning rate of $1 \times 10^{-5}$, a per-device batch size of 8. Training is configured for up to 5 epochs with a linear learning-rate scheduler and a warmup ratio of 0.02. Audio is resampled to 16~kHz, and the training objective is applied only to the target transcription tokens, excluding prompt tokens and padding positions.

\section{Result Details}
\subsection{Ablation Studies on TTS Adaptation Strategies}
\label{app:tts_ablation}

To investigate the effectiveness of the proposed TTS adaptation strategies, we conduct two additional ablation studies on CosyVoice2 and MaskGCT. Both experiments are evaluated using Gemini 2.5 Pro as an LLM-as-a-Judge, following the five-dimensional perceptual evaluation protocol. All scores range from 1 to 5, with higher values indicating better performance.

\subsubsection{Effect of Instruction Conditioning in CosyVoice2.}
We first examine the effect of instruction design in the second-stage adaptation of CosyVoice2. Both configurations use the same training dataset and differ only in their instruction-conditioning strategies. The single-instruction variant employs a unified natural-language instruction, whereas the dual-instruction variant distinguishes fluent and stuttered speaking states through separate natural-language instructions.

As shown in Table~\ref{tab:tts_adaptation_ablation}(a), dual-instruction conditioning achieves higher scores in four of the five evaluation dimensions. In particular, Perceptual Effect increases from 3.45 to 3.51, while Overall Expression improves from 3.79 to 3.90. Naturalness, Accuracy and Quality remain comparable between the two configurations. These results indicate that explicitly distinguishing fluent and stuttered speaking states is beneficial for instruction-based adaptation, particularly in terms of overall expressive quality.

\subsubsection{Effect of Masking Strategy in MaskGCT.}
We further investigate the masking strategy used in the third-stage adaptation of MaskGCT. The random-masking variant selects semantic-token regions without specifically targeting stuttering events, whereas the stuttering-region masking variant focuses reconstruction on regions aligned with annotated stuttering events.

As shown in Table~\ref{tab:tts_adaptation_ablation}(b), stuttering-region masking improves Accuracy from 3.84 to 3.91 and Perceptual Effect from 3.43 to 3.55, while maintaining comparable performance across the remaining perceptual dimensions. These results suggest that focusing the training objective on annotated stuttering regions is more effective for modeling local dysfluency events than random masking.

\subsection{Spectrogram Visualization of Stuttering Events}
\label{sec:stutter_visualization}

\begin{figure*}[t]
    \centering
    \caption{Representative spectrograms of four synthesized stuttering
    event types.}
    \includegraphics[width=0.96\textwidth]{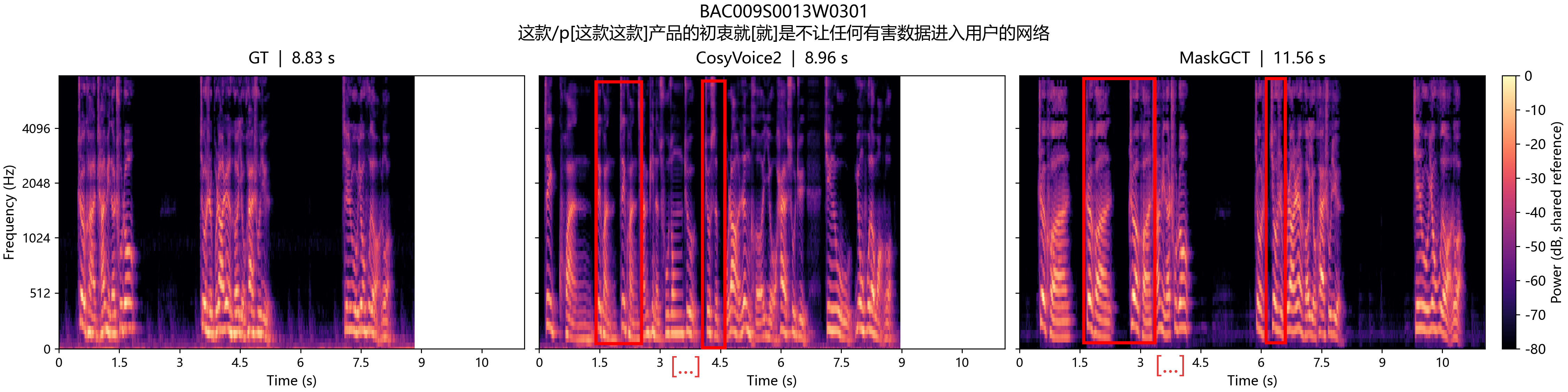}
    \par\vspace{0.2em}
    {\small (a) Word/phrase repetition (\texttt{[...]})}
    \vspace{0.7em}
    \includegraphics[width=0.96\textwidth]{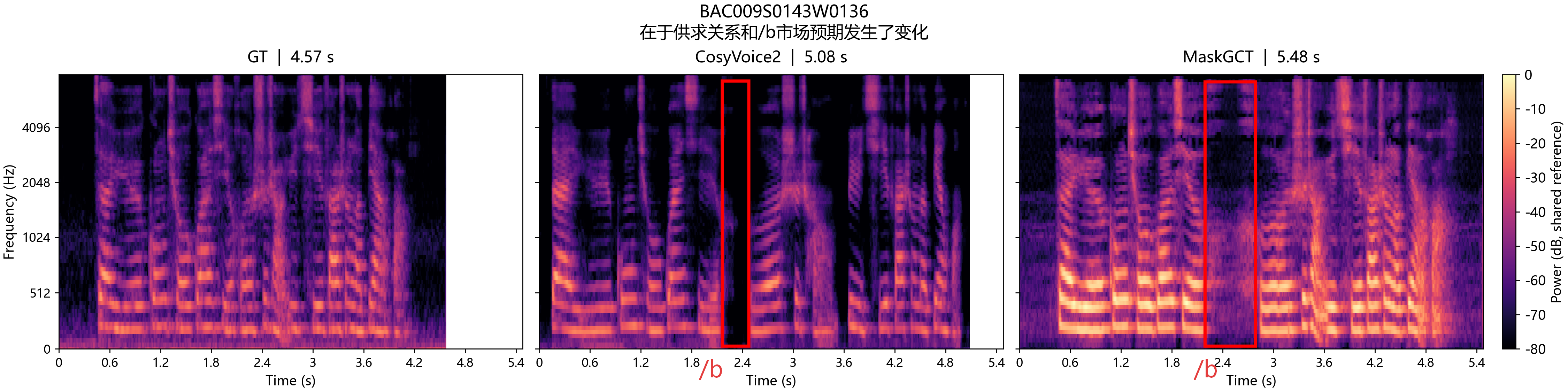}
    \par\vspace{0.2em}
    {\small (b) Block (\texttt{/b})}
    \vspace{0.7em}
    \includegraphics[width=0.96\textwidth]{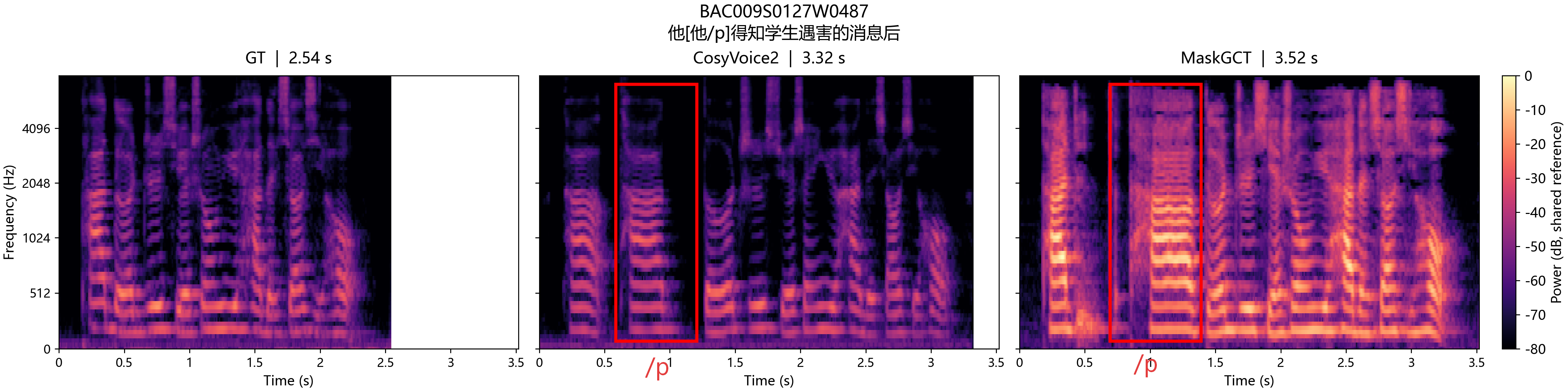}
    \par\vspace{0.2em}
    {\small (c) Prolongation (\texttt{/p})}
    \vspace{0.7em}
    \includegraphics[width=0.96\textwidth]{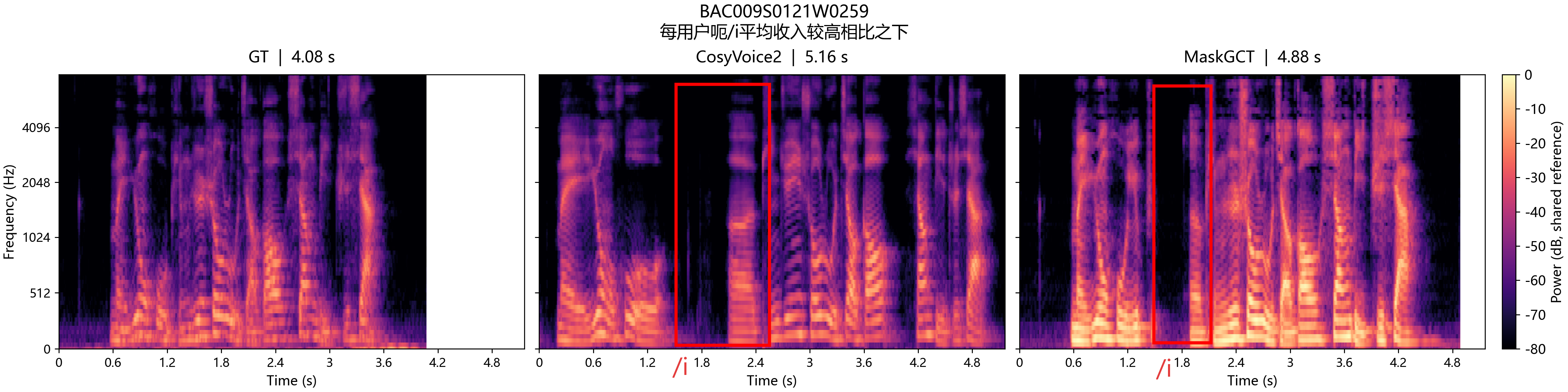}
    \par\vspace{0.2em}
    {\small (d) Interjection (\texttt{/i})}
    \label{fig:stutter_spectrograms}
\end{figure*}

To qualitatively illustrate the acoustic characteristics of the synthesized
stuttering events, Figure~\ref{fig:stutter_spectrograms} presents representative
spectrograms of four event types: word/phrase repetition, block, prolongation,
and interjection.

The four event types exhibit different acoustic and temporal patterns. Word/phrase repetition produces repeated acoustic structures corresponding to the repeated linguistic unit. Blocks introduce a temporary interruption in the speech signal, while prolongations extend the duration of a local speech component. Interjections introduce an additional hesitation or filler-like segment into the utterance.

Sound repetition (\texttt{/r}) is not separately visualized. It represents a finer-grained form of repetition, typically involving a short sub-character or phonetic unit. Because these events are usually brief, their repetition patterns are less visually apparent in a spectrogram than the complete word/phrase repetitions shown above.

\subsection{Additional Domain-Gap and Diversity Analysis}
\label{sec:appendix_gap_diversity}

To further support the domain-gap--diversity analysis in Section~5.2, we provide detailed speaker-wise distribution reductions and representation-diversity statistics.

\paragraph{Speaker-Wise Domain-Gap Reduction.}
Figure~\ref{fig:appendix_domain_reduction} presents the speaker-wise reductions in FSD and MMD$^2$, computed as the real--synthetic distance of CosyVoice2 minus that of MaskGCT. Positive values therefore indicate a smaller
domain gap for MaskGCT.

Both metrics show positive reductions for 43 of the 44 speakers, with speaker 0059 being the only exception. The magnitude of the reduction varies across speakers, but the overall pattern is highly consistent. This provides additional evidence that MaskGCT's smaller real--synthetic domain gap is not driven by a few
individual speakers. The results are consistent with its localized reconstruction strategy, which preserves most of the original acoustic context outside the
target stuttering regions.

\paragraph{Complementary Diversity Gain.}
Table~\ref{tab:appendix_diversity_detail} provides the complete representation-diversity statistics for the real AS-70 training set and its combinations with synthetic speech. Both synthesis systems increase the Vendi Score and covariance effective rank relative to real data alone, demonstrating
additional diversity in the combined representations.

CosyVoice2 produces a larger absolute Vendi Score gain than MaskGCT (0.3046 vs.\ 0.1343), corresponding to relative increases of 7.37\% and 3.73\%, respectively. A similar, though smaller, advantage is observed in covariance effective rank, with relative increases of 15.73\% and 14.48\%.

These findings provide quantitative support for the complementary properties of the two synthesis strategies: MaskGCT maintains greater target-domain proximity, while CosyVoice2 introduces more additional representation diversity when combined with real speech. Together, these measurements support the domain-gap--diversity interpretation of the downstream ASR results.

Importantly, the reported relative gains are computed as the equally weighted mean of speaker-wise relative increases, rather than the percentage change between two macro-averaged values. Specifically, we first calculate the relative increase for each speaker and then average these increases across all speakers.

\subsection{Event-Wise ASR Recognition Analysis}
\label{sec:event_wise_asr}

To further examine how synthetic stuttering data affect ASR performance around different dysfluency types, we report event-wise strict recognition accuracy for the five stuttering events. Due to space constraints, it is not practical to include the complete event-wise results for all training settings and both synthesis systems. We therefore use \textbf{AS70-Real} and \textbf{MaskGCT-Combined} as representative settings to illustrate the effect of synthetic-data augmentation. AS70-Real uses only the original AS-70 training data, whereas MaskGCT-Combined jointly uses the original AS-70 data and its MaskGCT-generated synthetic counterpart.

As shown in Table~\ref{tab:appendix_five_event_accuracy}, incorporating MaskGCT-generated synthetic AS-70 generally improves event-region recognition for a broad range of ASR systems, although the magnitude of the gain varies across architectures and event types. In particular, Wav2Vec2-CTC benefits consistently across all five stuttering categories, while Paraformer and several Whisper and Qwen3-ASR variants also show improvements on multiple event types. The effect is therefore not limited to a single dysfluency category or recognition paradigm.

At the same time, several models exhibit small degradations on individual event types, indicating that synthetic augmentation does not affect all architectures or dysfluencies uniformly. Nevertheless, the overall accuracy improves for most of the evaluated ASR systems under the MaskGCT-Combined setting. These event-wise results further support the main-text observation that synthetic stuttering speech provides useful complementary supervision, particularly for recognizing linguistic content affected by dysfluency.

\end{document}